\documentclass{amsart}

\usepackage{amsmath}

\usepackage{amssymb}

 \usepackage{mathptmx}      

\usepackage{graphicx}

\usepackage{amsfonts}

\usepackage{bbm,dsfont}

\newcommand{\tr}{\textrm{tr}} 

\newcommand{\sfq}{\mathsf{Q}}

\begin{document}

\title{Unsharp Quantum Reality}

\author{Paul Busch} 
\address{Department of Mathematics, University of York, UK}
\email {pb516@york.ac.uk}
\author{Gregg Jaeger} 
\address{Communication and Measurement Laboratory, Department of Electrical and  Computer Engineering
             	and Division of Natural Science and Mathematics, Boston University, Boston, MA} 
              \email{jaeger@bu.edu}

\date{July 2010}

\maketitle

\begin{abstract} 
The positive operator (valued) measures (POMs) allow one to generalize the notion of observable 
beyond the traditional one based on projection valued measures (PVMs). Here, we argue that this 
generalized conception of observable enables a consistent notion of {\em unsharp reality} and with it
an adequate concept of joint properties. A sharp or unsharp property manifests itself as an element 
of sharp or unsharp reality by its tendency to become actual or to actualize a specific measurement 
outcome. This actualization tendency---or {\em potentiality}---of a property 
is quantified by the associated quantum probability. The resulting single-case interpretation 
of probability as a degree of reality will be explained in detail and its role in addressing the tensions 
between quantum and classical accounts of the physical world will be elucidated. It will be shown that 
potentiality can be viewed as a causal agency that evolves in a well-defined way.



\end{abstract}

\section{Introduction}

The advent of Quantum Mechanics\footnote{We will use the expression
{\em Quantum Mechanics} as referring directly to the formal theory describing quantum mechanics, 
in contradistinction to {\em quantum mechanics} which refers to the physical phenomena
the theory is intended to describe.} has created a puzzling and, in fact, unique situation in physics. On the one hand 
this theory turned out to be enormously successful as a framework for the explanation of the behavior of matter at 
all scales---from the level of elementary particles, atoms and molecules to the world of macroscopic and 
even cosmic phenomena; on the other hand, researchers have not been able, in more than eight decades, 
to reach a consensus on its interpretation. There is not even an agreement over the status and
significance of such strange, uniquely quantum features as incommensurability, indeterminacy, indeterminism, and non-locality.
Nor is it considered established whether the so-called quantum measurement problem or the classical limit 
problem are actually {\em conceptual problems} or are instead artifacts of some particular choices of interpretation.

That quantum mechanics bears profound consequences for our physical world picture  became evident from
the early beginnings, as witnessed, for example, by Albert Einstein, who expressed his reservations against the
irreducibly indeterministic character of this theory in a letter to Max Born from 1926:
\begin{quote}
``Quantum mechanics is very worthy of regard. But an inner voice tells me that this is not yet the right track. 
The theory yields much, but it hardly brings us closer to the Old One's secrets. I, in any case, am convinced 
that He does not play dice.''
\end{quote}
By contrast, with his discovery of the uncertainty relations in 1927 \cite{Heis27}, Werner Heisenberg felt 
forced to conclude that through these relations quantum mechanics had definitively invalidated the principle of 
causality.

Nearly four decades later, at a time when the research area of Foundations of Quantum Mechanics was on its way to becoming
a research field in its own right, Richard Feynman espoused his views towards it with these words (in his book  
of 1965, {\em The Character of Physical Law}) \cite[Chap.~6]{Feynman1965}:
\begin{quote}
``\ldots I think I can safely say that nobody understands quantum mechanics. \ldots Do not keep saying to yourself, 
if you can possibly avoid it, `But how can it be like that?' because you will get `down the drain', into a blind 
alley from which nobody has yet escaped. Nobody knows how it can be like that."
\end{quote}
It is possible that statements like  this  led to the attribution of the slogan ``Shut up and calculate!" to Feynman where it is actually
claimed by David Mermin \cite{Mermin2004}, who once used it as a description  of the Copenhagen
interpretation of Quantum Mechanics and its perceived instrumentalist flavor.

These quotations capture the tension between two opposing philosophical positions: scientific realism versus 
instrumentalist empiricism. On the one hand, Einstein's concern was to uphold a world view based on
what is commonly referred to as ``local realism,'' in which probability plays a primarily epistemic role, whereas Heisenberg 
was prepared to  accept quantum indeterminacy and probability as primarily ontic, that is, as essential features of the physical 
world. On the other hand, there is still a strong presence of the view that Quantum Mechanics is nothing more than a formalism 
for the calculation of measurement statistics.

Many physicists now adopt a pragmatic double approach: they practice a realist outlook 
for the purposes of heuristic explorations of new models and the discussion of experiments, using intuitive 
pictures of individual (sub)atomic objects; but when challenged, they only admit to the minimal probabilistic or 
statistical interpretation of Quantum Mechanics. This conflicted attitude has similarly been noted by D'Espagnat \cite{DESP}.

It seems to us that a more coherent and productive approach would be to investigate systematically all possible 
variants of realist interpretations of Quantum Mechanics, including those in which probabilities are not
essentially epistemic. On a realist interpretation of Quantum Mechanics as a complete theory, the referent of quantum mechanical 
propositions is the individual system. This would not only recognize the possibility that such a philosophically realist 
interpretation could in the end enable the best description of the physical world; it also has the potential benefit of providing us 
with guidance in developing new, appropriately adapted intuitions about microphysical objects. Nonetheless, here we do 
not venture to embark on a comprehensive survey and critique of all known realistic interpretations. 
Instead, we identify a key element of such interpretations and explore a particular way of formalizing it.
A primary role of any realist interpretation is to provide a rule that determines, for every 
state, which physical quantities have definite values in that state, thus representing {\em ``elements of reality"}
or {\em actual properties} pertaining to the quantum system under investigation.

In the present paper we put forth an interpretational point of view that has not yet been much considered but 
that seems to us worthy of further exploration---the concept of {\em unsharp quantum reality}.  
Explicating this notion requires us to review the notion of elements of (sharp) reality (Subsection \ref{subsec:elem-real}) and the
issue of quantum indeterminacy  and its implications  (Subsections \ref{subsec:vague} and \ref{subsec:indet}). 
We then proceed to describe the attribution of sharp or
unsharp properties to a physical system, and the associated assignment of degrees of reality
(Subsections  \ref{subsec:degree} and \ref{subsec:unsharp}). We show that the elements of unsharp reality have the
potential of reconciling some if not all contrasting aspects of the quantum and classical descriptions of physical systems 
(Section \ref{sec:quclass}). However, along the way it will be seen that due to the notorious quantum measurement
problem, such reconciliation will require, under the proposed interpretation, a stochastic modification of the unitary time 
evolution postulate for quantum dynamics. Finally (Section \ref{sec:uqr}), we summarize the individual interpretation of Quantum Mechanics
resulting from the notion of unsharply real properties, showing how it brings more aspects of a quantum system into the realm
of the ``speakable" in the sense of John Bell's notion.

We adopt the standard Hilbert space formalism of Quantum Mechanics. States are represented by
{\em density operators}  $\rho$ (positive operators of unit trace), the pure states among them being given by
the rank one projections $P=P_\psi$, or one of their representative (unit) vectors, $\psi$. Observables are associated with positive 
operator measures (POMs). The traditional notion of an observable as a self-adjoint operator is included in the form
of projection valued measures (spectral measures of the self-adjoint operators); such observables
will be referred to as {\em sharp observables}, and all other observables will be called {\em unsharp observables}. 

Many of the ideas laid out in this work are inspired by discussions with Peter Mittelstaedt as well as his lecture 
courses and philosophical work. One of us (PB) owes Peter Mittelstaedt a great debt of gratitude for his 
generous guidance and mentorship over many years. It is a pleasure to have the opportunity to dedicate this 
work to him in honor of his 80$^{th}$ birthday.

\section{Potentiality: From Quantum Indeterminacy to Indeterminism}\label{sec:indet}

\subsection{Elements of reality}\label{subsec:elem-real}

On a minimal probabilistic interpretation, Quantum Mechanics is understood as a formalism for the calculation
of probabilities that correspond to predicted frequencies of the outcomes of measurements performed on identically prepared 
systems. Thus if the state resulting from the preparation is given by operator $\rho$ and the observable being
measured is a POM $E$ on a $\sigma$-algebra $\Sigma$ of subsets of the value space $\Omega$,  the associated
probability measure is 
\begin{equation}
\Sigma\ni X\mapsto p_\rho^E(X):={\rm tr}[\rho E(X)]\in[0,1].
\end{equation}

In a realistic interpretation, one is first of all concerned with identifying the formal elements representing the
properties or associated propositions that may or may not pertain to the system. These elements are commonly taken to be 
the eigenvalues or more generally the spectral projections of the sharp observables. The orthocomplemented lattice of
orthogonal projections (or the associated closed subspaces, with the order being defined as inclusion) forms the basic
proposition structure for an individual quantum system. Maximal properties are given by the atoms of the lattice, that is, the rank-one
projections (or one-dimensional subspaces). 

There is a far-reaching analogy between the set $\Gamma_q$ of one-dimensional
projections and the phase space $\Gamma_c$ of a classical mechanical system, which, we believe, speaks strongly in favor
of the view that the referent of Quantum Mechanics is an individual system: first, both sets carry a natural symplectic structure, making them
phase spaces, and second, their points represent the ontic states, which correspond to situations of maximal 
information.\footnote{A realist view of wave functions or Hilbert space vectors as ontic states has already been forcefully advocated by 
Birkhoff and von Neumann in their seminal 1936 paper on quantum logic \cite{BvN1936}.}
From this perspective, a closed quantum mechanical system with its unitary Schr\"odinger dynamics can be viewed as a (generally 
infinite-dimensional) Hamiltonian dynamical system. As will  be reviewed in Subsection \ref{subsec:dequant}, the analogy between 
$\Gamma_c$ and $\Gamma_q$ extends also to the treatment of probabilities in that all quantum states are representable as 
probability distributions on $\Gamma_q$.

The generalized notion of a quantum observable as a POM entails an extension of the lattice of projections to the 
partially ordered, convex set of {\em effects} (the positive operators that may occur in the range of a POM). Projections
are interpreted as representing \emph{(sharp) properties} of a quantum system. Effects represent \emph{quantum events} that may occur as outcomes of a measurement; such an event will be referred to as {\em sharp} if the effect is a projection and otherwise as {\em unsharp}.

A subclass of effects of particular interest are the {\em regular} effects, characterized by the property that their spectrum extends
both above and below the value $\frac 12$. We will explain below that regular effects that are not projections may be 
taken to represent {\em unsharp properties}. Along the way we will indicate the analogies  and crucial
differences between the concept of unsharp property and the notions of fuzzy sets and vague propositions.
The effects in the range of an observable are typically associated with fuzzy rather than sharp or crisp subsets of values.

We will formulate a general rule for the attribution of a degree of reality to an effect and the associated sharp or fuzzy value of 
an observable. In preparation we provide next an argument for the orthodox property attribution rule,  also known as the 
{\em eigenstate--eigenvalue link}.  In particular, we will argue that this rule, being based on an operational criterion,
ensures the {\em empirical} reality of a property, in contrast to the crypto-reality (or fairy-tale reality, to use a phrase coined by 
G\"unther Ludwig) of hidden variables or of properties and values attributed according to the various modal or histories 
interpretations, including the relative-state interpretation.\footnote{For example, in the relative-state interpretation, all possible 
measurement outcomes are considered equally real for a fictitious, non-existent ``observer" outside the Universe who views
the universal state vector, whereas for each observers inside the Universe it is always one particular value of
the measured observable that is real; and it is admitted that this reality can be described 
by the state vector collapse and the eigenvalue--eigenstate link---but only {\em for all practical purposes} (``FAPP") .}

A {\em sufficient} condition for a physical quantity to have a definite value, or for a property to be actual, is that the value or
property can be ascertained without changing the system. This is the famous Einstein--Podolsky--Rosen criterion for
a physical quantity to correspond to an {\em element of reality}. Using the apparatus of the quantum theory of measurement, one 
can show \cite{QTM} that there is a distinguished class of measurement operations that do allow just such procedures 
of determining the value of an observable without changing the state of the system. These operations are the so-called
{\em L\"uders operations}, which are restrictions of linear, positive, trace-non-increasing maps on the linear space of
self-adjoint trace-class operators. In fact, let $P$ be a projection, then the associated L\"uders operation is defined as
\begin{equation}\label{eqn:Luders1}
\rho\mapsto \phi_L^P(\rho):=P\rho P.
\end{equation}
A state $\rho$ remains unchanged exactly when it is an eigenstate of $P$, in the sense that $P\rho=\rho(=\rho P)$.
Thus, if the system's state is an eigenstate of a given observable, that observable can be said to correspond to an 
element of reality. This shows that the EPR condition is naturally implemented in Quantum Mechanics if one posits
the rule that a property is actual if the system's state is an eigenstate of the corresponding projection or observable.

In considering the question of what might constitute a {\em necessary} condition for a property to be actual, one may 
arrive at an answer by reflecting on the usual content of the terms ``real" or ``actual." A {\em thing} (Latin: {\em res}) 
has the capacity to make itself felt under appropriate conditions, it has the capacity to {\em act} on other elements of 
its environment. Hence the actuality, that is, {\em empirical} reality, of a property of an object 
entails that it has the ccapacity of influencing other objects---in particular, measurement devices---in a way that is 
characteristic of that property.\footnote{The sufficient and necessary conditions for the reality of physical objects or properties
proposed here are in accordance with an account of conditions of objective experience in the spirit of Kant that is based on 
the categories of substance, causality, and interaction; as Peter Mittelstaedt and his collaborator Ingeborg Strohmeyer 
have shown, this account can be cogently phrased in a way that encompasses objects of quantum-physical 
experience \cite{Mitt76,Stroh87,Mitt94}.}

Here, the properties of concern are physical magnitudes of a quantum system. If a property is absent, the system's action 
or behavior will be different from that when it is present. Applied to the context of a measurement, in which an observer induces 
an interaction between the system and part of its environment (a measurement apparatus), this means that if a property
is actual---that is, an observable has a definite value---then its measurement exhibits this value or property unambiguously and 
(hence) with certainty. This condition---which, incidentally, is routinely being used as a {\em calibration condition} for measuring 
instruments---is taken as
 the defining requirement for a measurement scheme to qualify as a measurement of a given observable  \cite{QTM}. 
Again, its implementation within the quantum theory of measurement is possible if a property's being actual is associated 
with the system being in a corresponding eigenstate.

In summary, the structure of the quantum theory of measurement, specifically the existence of L\"uders operations, 
suggests the adoption of the eigenvalue--eigenstate link as a necessary and sufficient criterion of {\em empirical}
reality in quantum mechanics. However,
the application of Quantum Mechanics to the description of measurement processes also leads directly into the
{\em quantum measurement problem}, or {\em objectification problem}, to which we will return shortly.

This problem is one of the main reasons for the continued debate about the interpretation of Quantum Mechanics.
Insofar as an interpretation entails its own specific rule for the attribution of definite properties, it is interesting to
note that effectively for each interpretation, its property assignment rule is uniquely determined by the quantum state 
and a particular observable that is always definite; this ``preferred" observable is specific to the interpretation at hand. 
This result, which we will not spell out here, is the content of the Bub--Clifton uniqueness theorem for (modal) interpretations of Quantum
Mechanics \cite{Bub97}. It highlights an important fact: any interpretation other than the one based on the eigenvalue--eigenstate link
faces the problem of explaining the definiteness of its particular  ``preferred" observable. 
The eigenstate--eigenvalue rule is based on the preferred observable being the identity operator, which could be trivially 
regarded as having a definite value.

\subsection{Vagueness and quantum properties}\label{subsec:vague}

We now provide a general consideration of the third possibility remaining beside the two considered above, that 
of a property being neither actual nor absent; the next subsection proceeds to a more formal discussion.

A property of an entity is typically thought of as {\em indeterminate} when there is no way, even in principle, of 
attributing it in pertinent contexts. In such situations, a corresponding vagueness of propositions regarding the 
properties of objects can arise, for example, when the concepts or definitions involved are either insufficiently 
clear or overly restrictive. However, quantum indeterminacy differs from these forms of vagueness.

In classical physics, which is known to describe correctly macroscopic objects (to a hitherto practically
unlimited accuracy) when quantum effects are imperceptible, all magnitudes have a 
definite value at any given time that can in principle be simultaneously known with certainty at that moment 
by their measurement, even though they may not always then be predicted with certainty; for example, when 
the classical evolution is chaotic, or may be difficult to measure. Hence, vagueness is considered alien to 
classical physics. 

By contrast, vague descriptions of macroscopic systems readily arise in geography, in that the boundaries 
of, for example, a mountain  are often vague in some respects such as their precise extent in square meters. 
Indeed, any conceivable quantitative criterion that might be adopted as a convention for the specification 
of the boundary line of a mountain must be expected to turn out too restrictive to be applicable in some cases. 
That sort of vagueness is less a question regarding objective reality than one of a choice of concepts; in the 
above instance it arises when an overly restrictive concept corresponding
to the property of geographical location is imposed. 

In the case of microscopic entities to which Quantum Mechanics pertains, there is vagueness the origin of
which is substantially different. Vagueness in quantum mechanics appears endemic, arising directly from the 
indeterminate nature of quantum entities themselves rather than a choice of concepts within a flexible theoretical 
framework. Indeed, in quantum physics spatial location can be almost {\em entirely} indeterminate, such as 
when the momentum is specified with extremely high precision, say in the case of a free electron.
Unlike the situation of macroscopic entities with vague geographical characteristics, it is not the case 
that the concept {\em electron} is overdetermined in the sense that the criteria for it to have a unique spatial 
location---or for that matter, for it to be spatially dispersed---cannot be satisfied in principle; 
nor is the concept of electron underdetermined  in the sense that the definitions of its physical properties are 
insufficiently precise \cite{KripsP31}. 

Quantum mechanical position is typically indeterminate but can, in principle, be
measured precisely  (i.e., with arbitrary accuracy) and be determinate (to a high degree) immediately after 
measurement (but with the result that momentum is disturbed and indefinite, and vice-versa). However, at least one of 
the two quantities, position and momentum (and typically each of these), is also  limited in its determination. 
In the case of quantum systems, properties can be considered objectively indefinite and sets of propositions regarding 
them complementary to specific other sets of propositions, so that it becomes impossible to jointly attribute them. 
Thus, quantum mechanics involves a unique form of vagueness distinct from those considered before.

If a traditional quantum observable $A$ has an eigenvalue $a$, {\em i.e.} $A\psi= a\psi$, then when the
state {\em is} an eigenstate of $A$  a measurement of that observable will yield the value $a$ with certainty; 
the physical magnitude corresponding to the observable $A$ has a definite value if  the state of the system is 
in one of its eigenstates, under the standard interpretation. However, there are always some other observables 
that have no definite value. It is the linear structure of the Hilbert-space formulation of Quantum Mechanics that
is responsible for the indeterminateness of properties: Every pure state is a non-trivial linear combination of 
eigenvectors of observables with which the associated density operator (one-dimensional projection) does 
not commute, so that the values of those observables are indefinite. The indeterminacy relation for state 
preparations governs the (in)definiteness of joint properties of quantum systems.

\subsection{Indeterminacy}\label{subsec:indet}

The novel feature in the description of physical reality brought about with the advent of quantum
mechanics is thus the fact that physical properties will not  in general be either actual or absent but 
{\em indefinite} or {\em indeterminate}.\footnote{In \cite{QTM} the term
{\em objective} is introduced to characterize a property that is either actual or absent in a general quantum state represented
by a density operator. In the present paper we consider mostly pure states and in this context use {\em actual}, {\em determinate} 
or {\em definite} as synonym to {\em objective}, and likewise for the negations.}
Thus it is impossible, according to Quantum Mechanics, to ascribe definite 
truth values to all propositions regarding all properties of a physical system at any one
time, for any of its quantum states. This quantum feature is acknowledged in all realistic interpretations
of Quantum Mechanics as a complete theory.

There has been a variety of approaches 
and attempts to demonstrate the inevitability of quantum indeterminacy, starting from different premises.
\begin{enumerate}
\item[(I1)] {\em Interference:} We adopt as the basis for our discussions an interpretation according to which 
an observable is assigned a definite value if, and only if, the system's 
state is an eigenstate of the observable (the eigenvalue--eigenstate link). On that interpretation there are
obviously many observables for every given state that do not have definite values---they are {\em indeterminate}.
This indeterminacy is exhibited in the presence of {\em interference terms} in the probability distribution of a suitable
test observable. For example, if $\phi_A$, $\phi_B$ are the two orthogonal states representing passage along
disjoint paths $A$ and $B$ in a split-beam experiment, and the system's state is $\psi=(\phi_A+\phi_B)/\sqrt 2$, 
the indeterminateness of the path observable  is made manifest by analyzing 
the probability distribution of an {\em interference observable} with eigenstates $\psi_\pm:=(\phi_A\pm\phi_B)/\sqrt 2$. 
Indeed, in a situation in which the system's path observable was definite---which is represented by either path 
eigenstate $\phi_A$ or $\phi_B$---the
probabilities for both outcomes $\pm$ of the interference observable would be 1/2 if there was maximal ignorance about
the actual path; whereas when the system's state is $\psi=\psi_+$, these probabilities are 1 and 0, respectively.

\item[(I2)] {\em Bell--Kochen--Specker theorem:} In a system represented by a Hilbert space of dimension greater than 2,
it is impossible to consistently assign {\em truth values} (1 for ``true", 0 for ``false")  to all elements of all complete 
orthogonal families of rank-1 projections in such a way that for each basis the value 1 is assigned just once. 
(An orthogonal family of projections is complete if the sum of all the projections is equal to the unit operator.) 
This result entails the impossibility  of {\em non-contextual} hidden-variable supplementations of Quantum Mechanics.

\item[(I3)] {\em Gleason's theorem:} Any (generalized) probability measure on the lattice of orthogonal projections of 
a complex Hilbert space of dimension greater than 2 is of the well-known form $P\mapsto p_\rho(P)={\rm tr}[\rho P]$, 
where $\rho$ is some positive operator of trace equal to 1 (a density operator). This entails that there are no dispersion-free
probability measures on the complete lattice of all projections. (Hence the Bell--Kochen--Specker theorem follows.)

This last statement can be extended to include unsharp quantum events represented by effects: any generalized 
probability measure on the set of effects is of the usual quantum mechanical form, $E\mapsto p_\rho(E)=
{\rm tr}[\rho E]$, with $\rho$ being some density operator. Owing to the stronger premise, this result is found to hold
also for Hilbert spaces of dimension two.

\item[(I4)] {\em Bell's theorem:} In the description of EPR experiments involving entangled, spatially separated systems, 
Bell's inequalities follow from the conjunction of assumptions of ``realism'' and ``locality.'' As Bell's inequalities are violated 
in the quantum mechanical description of these experiments, this shows that local realistic hidden-variable models and 
their ensuing value attributions are not compatible with Quantum Mechanics. EPR--Bell experiments have confirmed the
quantum mechanical predictions.
\end{enumerate}

The generally accepted upshot is that Quantum Mechanics does not admit global, {\em non-contextual} value attributions
to {\em all} physical quantities at once. It has been shown (for example, by Bell) that global value assignments are
possible provided they depend on the measurement, i.e., they are {\em measurement-contextual}; that is, the same 
projection may take a different value depending on the observable in whose range it occurs. The best elaborated and 
known example of a viable hidden-variables approach to quantum mechanics is Bohm's theory, which gives
the position variable of a quantum system the special status of a definite quantity. The Schr\"odinger representation of 
the state vector is considered to take the role of a guidance field. On that approach, in a two-slit experiment, each 
quantum system is ascribed a definite path that passes either through slit A or B, but it remains unknowable as a matter 
of principle through which path the system passes. In fact, the quantum system comprises the particle aspect, represented 
by the position and path, {\em and} the guidance field, and the latter propagates through both slits and acts non-locally.

\subsection{Degree of reality, actualization, and measurement}\label{subsec:degree}

We noted above that, as an element of empirical reality, an actual property has the capacity to act, 
to actualize an indicative measurement outcome if a measurement is performed. 
By contrast,  when a property is absent it has no capacity to act. 
We propose the idea of an interpolation between the two extremes of full actuality and absence of a property.
Hence, we propose to say that an indeterminate property, being neither fully real nor completely absent, 
possesses a limited degree of actuality, or reality. This is justified insofar as
an indeterminate property has a quantifiable, limited 
capacity---{\em potentiality}\footnote{This term was introduced by W.~Heisenberg to 
express the tendency of quantum events to actualize in a measurement \cite{H1,H2}.  
A similar idea was expressed by K.~Popper \cite{P1,P2}, who used the term {\em propensity} 
to refer to the objective single-case probabilities of indeterminate quantum events.}---to cause 
an indicative measurement outcome. A quantitative measure of this potentiality and degree of reality
is given by the quantum mechanical probability, which provides the likelihood for an individual outcome 
to occur in the event of measurement. 
We show now how this interpretation of the quantum state $\psi$ in terms of single-case probabilities
is supported by the quantum theory of measurement.

Measurement is commonly understood as a process in which a physical system (the {\em object} system) is 
made to interact with a {\em probe} or {\em apparatus} system in such a way that the apparatus ends up 
indicating the value of a particular physical quantity pertaining to the system under investigation. This indication 
is modeled by means of a {\em pointer observable} assuming an actual value that corresponds to a value of the system 
quantity of interest.  

As also noted above, a minimal condition for a measurement process to qualify as a measurement of a given observable is
the {\em calibration condition}: if the system is in an eigenstate of that observable when measurement begins, then the 
corresponding eigenvalue will be indicated with certainty. This means that the evolution of system plus apparatus is such 
that the apparatus state at the end of the joint evolution 
is the associated pointer eigenstate. If the state is a superposition of eigenstates of the observable being measured, then 
according to the minimal interpretation, the probability of the occurrence of a particular pointer value is given by the squared 
modulus of the amplitude of the corresponding object eigenstate. In the framework of the quantum theory of measurement, 
this result is in fact a consequence of the calibration condition \cite{QTM}. 

Adapting arguments initially developed in the 
context of the relative-state interpretation of Quantum Mechanics, Peter Mittelstaedt has shown that the quantum mechanical 
probability emerges as an (approximately) actual property of a large  ensemble of identically prepared systems represented 
by the same state \cite{Mitt98}. Specifically, it is shown
that if the same measurement of a given quantity (with a discrete spectrum, for simplicity) is carried out on all members of an
ensemble of equally prepared object systems, each in state $\psi$, the ensemble of object-plus-apparatus systems ends up
in a state in which the {\em frequency operator} associated with each value of the apparatus pointer observables has assumed an actual
value, which is identical to the quantum mechanical probability of the measured observable in the state $\psi$.
Moreover, if the measurement is {\em repeatable}, it follows also that the frequency operators associated with the values of the 
measured observable will have assumed the associated quantum mechanical probabilities as their (approximately) actual values.
The quantum mechanical probability postulate is thus found to be deducible from the probability-free
eigenvalue--eigenstate rule, which specifies the actual properties of a quantum system in any given state.

However, the entity that carries the probability value as an (approximately) actual property is an {\em ensemble}
of equally prepared quantum systems. Since this system is composed of independent subsystems, the question remains how 
one could then explain that such independent objects act together to cause the emergence of an approximately actual 
eigenvalue of the respective frequency operators of the ensemble. After all, the description of the 
ensemble as a compound system is hardly more than a conceptual construct. The answer is therefore to be sought in the 
fact that each constituent has been treated---{\em prepared}---in an identical way.  One may phrase an explanation as follows: 
Through its preparation, each individual system is constrained to ``respond" to a measurement  
so as to induce a specific outcome in accordance with the probability law specified by its quantum state. 
Each possible value of the observable to be measured has a limited degree of reality, quantified by the 
associated quantum probability. Accordingly, the individual has a limited capacity to cause the actualization of each 
of the corresponding pointer values (and, in a repeatable measurement, the values of the observable).  On many repetitions
(or in many parallel runs) of the measurement, the degrees of reality of the different values are then manifested as the
frequencies of the outcomes. 

The notion of an indeterminate property carrying a limited degree of reality also allows one to think of quantum measurements
more along classical lines as revealing what is actually the case: the fact that a particular outcome did occur entails that
the corresponding property was not completely absent. It is also appropriate to think of  an indeterminate
property as an element of {\em unsharp reality} in the following sense. If a property $P$ is indeterminate, then so is its 
complement $P^\perp=I-P$. Thus, both $P$ and $P^\perp$ have a nonzero degree of reality, they {\em coexist}, to a 
nonzero degree of actuality, in the given state.
In this sense they are both simultaneously but ``unsharply" defined. This description seems to be in agreement with Bohr's
account of the uncertainty relation: in a quantum state given by (say) a Gaussian wave function, the position and momentum
of the quanton are, according to Bohr, both defined with a latitude. Bohr uses the phrase ``unsharply defined individual" to characterize
this situation. In such a Gaussian state, both position and momentum are indeterminate, hence they correspond to elements of
unsharp reality. One can say more precisely that each spectral projection of position and momentum associated with a finite interval
has a degree of reality different from zero and different from one.

This ``unsharp reality" interpretation of quantum indeterminacy accounts for such  intriguing phenomena as the
emergence of an interference pattern in a two-slit 
experiment in which quantons are sent through the slitted diaphragm one by one, each giving a light spot in an apparently 
random location on the capture screen: what appears as random behavior at the level of an individual member of an 
ensemble is found to be guided by the (approximate) actuality of the path and interference properties of the ensemble. 
This account can be supplemented with one that reduces this ensemble behavior to that of the individual systems: 
The remarkable collective behavior of many, equally prepared, quantons can be explained by noting that the interference 
pattern produced by the ensemble is a manifestation of the potentiality inherent in, and quantified by, the state of the 
individual quanton.

Finally, it is a consequence of quantum measurement theory that quantum {\em indeterminacy} entails the 
{\em indeterminism} of measurement outcomes: if a property has no definite value, all a measurement can 
do is induce the random occurrence of one of the possible outcomes; that is, the individual outcome is 
not necessitated by any identifiable deterministic cause, yet this individual event is governed by 
{\em probabilistic causality}---the probability law commonly called Born rule, which attributes to it a 
well-defined, quantified potentiality, {\em qua} degree of reality. 
This conclusion rests on the premise  that measurements and the occurrence of their outcomes, being 
physical processes, are correctly described and explained  by Quantum Mechanics itself. What is at stake 
here is the {\em semantic consistency} of the theory; this is an issue of central  concern in Mittelstaedt's 
work \cite{Mitt98} that is called into question by the problem of objectification (see Sec.~\ref{sec:quclass}).

\subsection{From sharp to unsharp properties}\label{subsec:unsharp}

The idea of unsharp quantum reality was conceived originally with the aim of interpreting {\em unsharp observables}, that is, 
POMs that are not projection valued. It is therefore important (though perhaps at first surprising) to observe that the notion of 
unsharp reality is also applicable to sharp observables  (projection valued POMs). In fact, up to this point our consideration 
of these concepts mostly referred to sharp observables.

It was in the context of a discussion of the EPR--Bell experiment that the Einstein--Podolsky--Rosen condition of elements of reality 
was relaxed to a condition regarding elements of {\em approximate} reality  \cite{Busch85}. As the spin or polarization observables 
of the entangled particles in an EPR experiment are measured with 
progressively more limited accuracy, there is a corresponding progressive degradation of the degree of Bell inequality violation 
due to the appearance of additional terms reflecting the inaccuracy; the violation of the Bell inequality becomes unobservable above 
a certain degree of inaccuracy.

This consideration illustrates two facts: (1) the notion of elements of reality is robust in the sense that reality in the sharp sense can be
arbitrarily well approximated by measurements of appropriate unsharp observables with decreasing degrees of unsharpness;
(2) quantum observables with sufficient degrees of unsharpness may display classical features, such as the satisfaction
of Bell inequalities.

The {\em sufficient condition} for elements of reality given by Einstein, Podolsky and Rosen can be adapted to the realm of effects,
by way of a relaxation of its formalization for a sharp property given above. A good candidate of a measurement operation generalizing the 
L\"uders operation (\ref{eqn:Luders1}) for an effect $E$ is the following:
\begin{equation}
\phi_L^E:\rho\mapsto \phi_L^E(\rho)=E^{1/2}\rho E^{1/2}.
\end{equation}
It has been shown \cite{Busch86} that if ${\rm tr}[\rho E]\ge 1-\varepsilon>0$, then $\phi_L^E$ does not decrease this 
probability, so that it remains above $1-\varepsilon$,  and the state change, measured by the trace-norm distance, is small to the order of $\sqrt{\varepsilon}$.
This demonstrates the stability of the EPR reality criterion against small deviations from actualization, and one may say 
that approximately real properties can be ascertained almost with certainty and without significantly altering the system.
It is in this sense that an effect $E$ that is not a projection is still capable of being an element of approximate reality.

We propose also to adopt the {\em necessary} condition of the actuality of a property for effects in general: 
that an effect has the capacity to act, to cause the actualization of an indicative event if it is real, or actual. 
The quantum event represented by an effect is {\em indefinite}, or {\em indeterminate} in a pure state 
represented by $\psi$ if its probability is neither 0 nor 1. In such a situation, the event possesses the potentiality 
to actualize when measured, and it is appropriate say that the effect corresponds to an element of unsharp reality 
and is actual to the degree  given by the probability $\langle\psi|E\psi\rangle$.

An effect can thus be real to a degree that may vary between 0 (completely absent) and 1 (fully actual). We will speak, 
more specifically, of an effect being {\em approximately real} if its degree of reality is greater than $\frac 12$.
For an effect to represent a possible property, it should  be capable of being approximately actual {\em and} approximately 
absent in appropriate states, in the sense that its degree of reality is greater than $\frac 12$ in the first case and less than 
$\frac 12$ in the second case. Thus we recover the condition of an effect being regular, concisely described via
$\frac 12 I\not\le E\not\ge\frac 12I$. This allows us to regard a regular effect $E$ (together with its equally regular complement $E'=I-E$)
as  an {\em unsharp property}, thus relaxing the notion of {\em sharp properties} that is restricted to projections. We note that the weakened 
EPR condition for elements of unsharp reality can be applied to sharp properties  as well, which will therefore correspond to 
elements of approximate reality if the system's state is a near-eigenstate.

In general, an effect may only allow limited degrees of reality (namely, when its spectrum does not extend to 1), 
and the degree of reality may not be enhanced by a 
measurement (as would be the case with a generalized L\"uders operation). In such cases the above {\em sufficient}
EPR reality condition will be of little use. But one may refer to the necessary condition for elements of reality 
introduced here: any effect has a capacity to actualize an indicative measurement outcome.
The set of all effects of a quantum system then comprises the totality
of the ways in which the system may act on its environment, specifically in  measurement-like interactions. At this interactional
level, and in the sense of our necessary reality criterion, an effect can thus be described as a sort of {\em relational} property, or
disposition. If, in addition, these interactions are of such a nature that the influence on the system is described by a (generalized)
L\"uders operation, this can be used to ascertain the (approximate) actuality of a {\em property} of the system 
whenever this operation does not (significantly) alter the system state: in accordance with the EPR criterion, it can be said that what 
can be observed and therefore predicted without disturbance must already have been (approximately) actual.

In order to elucidate further the nature of the unsharpness of an effect as a quantum event or property, we may note an
analogy between the notions of {\em effect} and {\em fuzzy set}: just as a fuzzy set is characterized by a kind of 
``smeared-out" characteristic (set indicator) function of the associated crisp set, every effect can be presented as a kind of 
weighted distribution of projections. For example, if $P_\pm$ are the spectral projections of the $z$-component of the
spin of a spin-$\frac 12$ system, then $E_\pm=\alpha_\pm P_++\beta_\pm P_-$ (with $0\le\alpha_\pm\,,\beta_\pm\le 1$
and $\alpha_++\alpha_-=\beta_++\beta_-=1$) are
effects which are generally not projections unless $\alpha_\pm=1-\beta_\pm\in\{0,1\}$. The effects $E_+,E_-$ constitute a
POM which represents  a {\em smeared spin observable}.

The converse is also true: every effect of this system can be expressed as a weighted sum of two orthogonal projections. 
Owing to the spectral theorem, this fact holds for any effect in a (complex) Hilbert space of arbitrary dimension. There are 
also possibilities of expressing an effect as a weighted sum of nonorthogonal projections: for example, in the above case 
of a 2-dimensional Hilbert space, if $E=\alpha P_++(1-\alpha)P_-$ is the spectral representation of $E$, with $0<\alpha<1$, 
then for any rank-1 projection $R$ there is a unique number $\beta\in (0,1)$ and a unique rank-1 projection $R'$ (which is 
not orthogonal to $R$ unless $R$ is $P_+$ or $P_-$) such that 
$E=\beta R+(1-\beta) R'$. (We leave the proof as an exercise.)

Another example is given by the effects in the range of an unsharp position observable of a particle in a line (say). 
Let $\sfq$ denote the
spectral measure of the self-adjoint position operator $Q$, then for any positive distribution function $e$ on $\mathbb{R}$, the map
$X\mapsto \sfq^e(X)$ defines a POM on the Borel subsets of $\mathbb{R}$, where the effects $\sfq^e(X)$ are given by
\begin{equation}
\sfq^e(X)=\int e(x)\sfq(X+x)dx=\int(\chi_X\star e)(x)\sfq(dx)
\end{equation}
This is a {\em smeared position observable}. The effects $\sfq^e(X)$ are weighted sums of the projections 
$\sfq(X+x)$, and their spectral representation expresses them as smeared indicator functions of the set $X$. 
It is in this sense that  also the values (or value sets) of an unsharp observable can often be interpreted as 
{\em fuzzy values} (or fuzzy sets). A similar interpretation also applies in the above discrete example.

These examples show that two complementary effects $E=E(X)$, $E'=E(\mathbb{R}\setminus X)$
are {\em overlapping} if they are not projections, in the specific sense that their spectral representations
have contributions from a common subset of projections. This feature can be algebraically described by the
simple ``overlap" relation $EE'> 0$, or $E> E^2$. This is in line with the original definition of an effect being sharp
exactly when it is a projection.

It is important to note that the nature of the fuzziness of quantum effects differs fundamentally from that of fuzzy sets, however.
In the latter case, the rule for the application of one of a set of alternative fuzzy sets is based on there being an underlying 
fine-grained level of actual reality. For example, the predicates ``tall,"``medium-sized," or ``small" applied to describe the size 
of a person are typically defined by a choice of functions $f_t(x)$, $f_m(x)$, $f_s(x)$ on the set of possible sizes $x$ of a 
person, measured in centimetres, with values in $[0,1]$, such that, say, $f_t(x)$ is defined as increasing continuously from 
the value 0 for $x\in[0,1.5m]$ to the value 1 for $x\ge 1.9m$ (and similar definitions applying to $f_m,f_s$). In the case of a
quantum effect, we have seen that such valuations functions over subsets of projections exist but, as we have seen, these 
projections are generally indeterminate and cannot be regarded as elements of reality.

The above examples of unsharp observables were obtained as smearings of some underlying sharp observable 
(spin or position). In these cases the unsharpness or fuzziness arises from measurement inaccuracies, which can in principle
be made smaller and smaller. Nonetheless, this situation is not generic in Quantum Mechanics. There are important observables that are
represented as POMs which do not derive as smearings of some sharp observable. Examples are phase space,
phase and time observables. These are appropriately characterized as {\em irreducibly unsharp} quantum
observables.

We will freely use the terminology of elements of unsharp, or approximate reality and degrees of reality to the  fuzzy values
of unsharp observables, just as it is common practice in the case of sharp observables, their spectral projections, and their
sharp values.

\section{Unsharp Quantum Reality and the Quantum--Classical Contrast}\label{sec:quclass}

We now address the question of the extent to which the notion of unsharp quantum reality could contribute to a
resolution of some of the most notorious conceptual problems of Quantum Mechanics. Here we focus our attention
on the following aspects of the quantum--classical contrast: quantization; dequantization; 
the classical limit problem; and the problem of objectification. We will consider each of these issues in turn and then show
how they seem to be interrelated.

\subsection{Quantization}
Quantization is a procedure for formulating the quantum theory of a type of physical system using a form of {\em correspondence}
with an analogous classical system. In the original heuristic version introduced by the pioneers of Quantum Mechanics,  quantization 
consisted of associating self-adjoint operators on a complex Hilbert as representations of quantum observables with
certain functions on the phase space of a classical system corresponding to classical observables; this correspondence was to translate
the Poisson bracket relations of canonically conjugate variables such as position and momentum into commutator relations of the
operator counterparts. In a more refined fashion, quantum observables are defined through their covariance
properties under the fundamental spacetime symmetry group, the Galilei or Poincar\'e group. The group parameters are labels
that describe the---classically characterized---locations and orientations of macroscopic preparation and measuring devices. 

The problem faced with quantization consists of the fact that, on the one hand, Quantum Mechanics is considered to supersede 
Classical Mechanics while, on the other hand, it is not fully emancipated from the latter in that classically described macroscopic 
apparatuses located in a classical background spacetime are presupposed in any quantization procedure.

There are various strategies for addressing or eliminating this problem. One approach is to take the dualistic 
view (and hope) that physical theories of both objects and spacetime structures come in quantum and classical 
versions, each with their own domains of validity separated by different scales. Then one need not require that 
classical physics be reduced to quantum physics (or vice-versa).
However, advocates of this position have (so far) failed to provide an explanation for any hierarchic or emergence relationship 
between the theories for the small and those for the large that could account for the fact that the constituents of macro-objects 
are systems obeying the laws of quantum theory---and thus provide an explanation of the border between the supposedly
separate domains of validity.

An alternative approach is pursued in the search for a theory of {\em quantum gravity}, which takes quantum theory 
and the quantum structure of spacetime to be fundamental (so that classical objects and classical spacetime would 
be emergent, large-scale structures). One recent version of this approach has grown out of the $C^*$-algebraic formulation 
of physical theories by taking seriously the notion that spacetime coordinates should be described as quantum variables, 
thus giving rise to a concept of {\em quantum spacetime} \cite{Fred-etal1995}.
First promising steps towards formulations of quantum field theories on quantum spacetime have been reported in 
\cite{Fred-etal2003}, for example. Insofar as spacetime coordinate observables turn out to be noncommuting, the natural 
framework for a description of spacetime measurements is that of the theory of unsharp observables, as this allows one 
to account for the simultaneous measurability of these noncommuting coordinates.

\subsection{Dequantization}\label{subsec:dequant}
The question whether quantum mechanics can be recast in terms of classical physical theories has been 
asked ever since the discovery of quantum mechanics. This is the dequantization problem, also known as the 
problem of finding a supplementation of Quantum Mechanics by means of {\em hidden variables}, that is, 
classical quantities that have definite values in all states. In Subsection \ref{subsec:indet} we
surveyed arguments showing that quantum mechanics cannot be described in terms of a non-contextual hidden 
variables theory. We have also recalled Bub's uniqueness theorem that elucidates the possible partial value 
attributions given one distinguished, definite observable.

Perhaps surprisingly at first, it turns out that there is a formulation of Quantum Mechanics as a classical probability 
theory which yields a universal attribution of fuzzy values, or elements of unsharp reality. This formulation, which is 
based on the set $\Gamma_q$ of Hilbert space rays as ontic quantum states as the ``classical" phase space, 
was first studied rigorously in 1974 by B.~Misra \cite{Misra1974} and elaborated fully as a probabilistic model of 
quantum mechanics by S.~Bugajski in the 1990s \cite{Bug}.  The space $\Gamma_q$ is naturally
equipped with a Borel structure via the metric induced by the trace norm. The convex set of all probability measures
on $\Gamma_q$ forms the classical state space $S_c$. There is an affine map $R$ from $S_c$ to the convex set 
$S_q$ of all quantum states (density operators) given by the association 
\begin{equation}\label{eqn:MB-map}
S_c\ni\mu
\mapsto R(\mu)=\int_{\Gamma_q}P \mu(dP)\equiv\rho_\mu\in S_q.
\end{equation}
The integral is defined in the sense that for every probability measure $\mu$ there is a density operator 
$\rho_\mu$ such that for any effect $E$, 
\begin{equation}
{\rm tr}[\rho_\mu E]=\int_{\Gamma_q}{\rm tr}[ PE ] \mu(dP)\equiv \int_{\Gamma_q}f_E(P)\mu(dP).
\end{equation}
This is the probability of the occurrence of the effect $E$ in the state $\rho_\mu$. 

We note that the  map $R$ thus defined is many-to-one and associates with any
density operator all probability distributions with which it can be decomposed.
Further, the correspondence $E\mapsto f_E=:R'(E)$ defines  the map $R'$ dual to $R$, 
which sends all quantum effects to classical effects. 

We find that Quantum Mechanics is thus presented as a classical theory with the set of effects restricted to
a certain subset of fuzzy sets. In fact, even if $E$ is a projection other than $O$ or $I$, the corresponding 
classical effect $f_E$ is a fuzzy set function assuming all values in the interval $[0,1]$. Correspondingly, 
the set of observables is restricted to certain {\em fuzzy random variables}. As an interesting historical 
comment we note that this observation has inspired the development of a  classical fuzzy probability 
theory, as documented in \cite{Bug} and works cited there.

The classical presentation of Quantum Mechanics via the maps $R,R'$ is comprehensive in that all 
quantum states and quantum effects have their classical counterparts. Moreover, it has been shown to 
be essentially unique by W.~Stulpe and one of the authors \cite{StulpeBusch2008}. This result provides 
formal support to the interpretation of indefinite quantum events in terms of degrees of reality 
defined by the quantum probabilities. Indeed, the formal probability expressions ${\rm tr}[ E P]$
are recovered here as measures of the fuzziness of the unsharp property represented by $E$ while the
quantum probabilities ${\rm tr}[\rho E]$ are averages of the fuzzy set function $f_E$, weighted by any 
probability measure $\mu$ associated with $\rho$ via $\rho=R(\mu)$.

Moreover, quantum fuzziness is seen to be a reflection of a geometric peculiarity of the quantum 
phase space $\Gamma_q$: if $E$ is taken to be a rank one projection, $E=P'$, then $f_E(P)=\tr[PP']$; 
this quantity can assume any value between 0 and 1 for any given $P$, by varying $P'$. In particular,  
$f_{P'}(P)=0$ only when the two projections are mutually orthogonal. This shows that two maximal 
information states, or quantum phase space points, are ``overlapping"  rather than disjoint, and it is 
well known that this is the reason for the fact that two non-orthogonal states cannot be unambiguously 
separated by any single-shot measurement. In other words, two nonorthogonal pure quantum states always 
coexist  in that the maximal properties they represent are fuzzy, giving rise to nonzero degrees of reality, 
$f_{P'}(P)=f_P(P')$.

This phenomenon is geometric in nature insofar as the overlap $f_{P'}(P)=\tr[P P']$ is related 
to the operator norm distance between $P$ and $P'$ via
\begin{equation}
\|P-P'\|^2=\bigl[1-\tr[P P']\bigr].
\end{equation}
By contrast, two distinct points in a classical physical phase space are always disjoint, the associated 
probability norm generates a discrete topology where two points have distance 1 if they are identical 
and 0 if they are different. Seen from this perspective, quantum mechanics
appears {\em more continuous} than classical mechanics.

With this consideration the `canonical classical presentation' of Quantum Mechanics due to Misra and 
Bugajski explains the impossibility of a universal non-contextual hidden-variable formulation of Quantum 
Mechanics: insofar as this presentation is essentially unique, it shows that the only possible classical 
formulation is one whose global value ascription is intrinsically fuzzy. We note that this presentation 
displays a form of {\em preparation contextuality}, in that  a change of preparation is not always reflected 
in a change of state---one and the same mixed quantum state $\rho$ admits an infinity of different 
preparations, as reflected by the multitude of mixtures $\mu$ of point measures mapped to the same
$\rho$ \cite{Spekkens05}. This is another way of reflecting the fact that the classical random variables
selected by the map $R'$  to represent quantum observables are a small subset of all formal classical 
quantities and they are thus insufficient  to separate all classical probability measures beyond the degree
needed to distinguish two different quantum states.

\subsection{Classical limit of Quantum Mechanics}\label{subsec:clqm}
The classical limit problem arises if Quantum Mechanics is taken to be a universal theory, encompassing 
classical physics. Intuitively, one would expect that the typically classical behavior of a macrosystem 
should emerge within a quantum mechanical description if relevant quantities are large compared to 
Planck's constant. The classical limit is commonly thought to be associated with the formal limit of 
Planck's constant approaching zero, $\hbar\to 0$. However, it takes considerable effort to carry out 
this limit in a mathematically satisfactory way \cite{Landsman07}. Moreover, it is not clear whether this 
simple process is capable of capturing the emergence of classical behavior {\em in operational terms}. 
In fact, there is more involved than taking the limit of a single parameter, as will be explained in the following example.

It is often said that coherent states constitute the best approximation to the states of a classical particle or a classical electromagnetic field. 
Yet, it seems that this statement in isolation makes little sense. In order to see, for example, whether the centre of mass degree of freedom
of a large system could be regarded as approximately classical, it is necessary to consider the conditions under which
this variable could be said to have (relatively) well-defined values that can be measured without disturbing this
degree of freedom. As a classical variable one would expect the centre of mass  to have a definite position at all times,
which would, according to classical deterministic reasoning, require its momentum to be equally definite.  

The only way these requirements can be approximately reconciled with quantum mechanics is by way of representing the position and 
momentum of the centre of mass simultaneously as marginals of a quantum mechanical phase space observable. The approximate 
definiteness of the values of these quantities is given if they can be considered as relatively sharply defined on a macroscopic scale, 
while their noninvasive measurability requires their degrees of unsharpness to be large in comparison to Planck's constant. It then
turns out that minimum uncertainty states are those that are least disturbed in such macroscopic, coarse-grained phase space measurements.
Thus the underlying phase space observable must be equipped with macroscopically large inaccuracies. At the same time, in order for the
values of these macroscopic pointer observables to be considered `point-like,' their unsharpness has to be 
small on the readout scale.  A more detailed technical account of these classicality conditions for the observation of a macroscopic 
trajectory can be found in \cite{OQP} together with further references to relevant studies.

We see that the theory of unsharp observables and their interpretation as elements of unsharp reality provides necessary key elements
in the modeling of approximately classical behavior within quantum mechanics. More generally, within the set of general observables,
pairs of observables may be jointly measurable without having to commute with each other. A sufficient degree of unsharpness will allow
two POMs to be jointly measurable. Two POMs are called jointly measurable if they are marginals of a third POM. 
Joint measurability of a triple or quadruple of spin-1/2 observables entails the validity of Bell's inequalities for the pair distributions 
among these observables. This explains the fact, mentioned earlier, that Bell's inequalities are satisfied in the case of sufficiently unsharp
variables. If this happens in the context of an EPR experiment, the usual argument against ``local realism'' no longer applies. One may say
that in the case of sufficiently unsharp measurements, Bell-locality can be maintained as a classical fiction.

So far, we have discussed operational aspects of the classical limit problem. The universality claim of Quantum Mechanics  also entails that
there should be an account of the time evolution of macroscopic objects and, moreover, of hybrid microscopic-macroscopic systems, including
measurement processes. Given this requirement, we are faced with the quantum measurement problem.

\subsection{The problem of objectification}

In the context of the quantum 
mechanical description of a measurement, quantum indefiniteness in conjunction with the linearity of (the unitary) time 
evolutions gives rise to the {\em measurement problem}, or {\em objectification problem}: if the system's 
initial state is a superposition of different eigenstates, then by the linearity of the unitary coupling map, the total system 
state will evolve into an {\em entangled} state, each component of which corresponds to a different pointer value. 
Thus, the conclusion that a definite pointer value has been realized is not warranted.
This is the problem of {\em pointer objectification}. 

In a realist interpretation it is noteworthy that discrete observables (for example, sharp yes-no measurements)
admit {\em repeatable} measurements, in which the coupling interaction leads to the measured property being strictly 
correlated with the pointer in the final total state. Thus, if a definite pointer value is indicated, one can conclude that the 
measured system observable possesses the definite value associated with this reading. However, the objectification 
problem extends to those cases in which the initial state is a superposition of eigenstates. Then, the post-measurement 
state of the total system is not an eigenstate of the measured observable, and hence the objectification of that object 
observable is not warranted according to the quantum mechanical description of a repeatable measurement.

These negative results have been made precise in a succession of {\em insolubility theorems} initiated by Wigner. The 
presently most general version and a brief history of these theorems is given in \cite{BuShim96} and \cite{Busch98}. The
assumptions of the insolubility theorem are (UD) the linear, unitary dynamics of quantum states,  (R) the reality criterion
given by the eigenvalue--eigenstate link, and (DO) the stipulation that a physical measurement process terminates 
with a definite outcome. These three requirements lead to a contradiction. One can eliminate the objectification problem
by adopting some modified form of the the reality criterion (R); in each there will be a specific way of accounting
for definite outcomes (DO). This has the advantage of leaving the formalism of Quantum Mechanics intact and untouched 
but---as made evident by the Bub--Clifton uniqueness theorem---it raises the problem of explaining the special status
of one distinguished observable. Moreover, as with the ``many-worlds" interpretation, it appears as though in these 
interpretations, Quantum Mechanics describes the world as watched ``from outside" by some non-physical entity. In other 
words, the value attribution rules of such interpretations seem to endow quantum properties only with a sort of crypto-reality. 

By contrast, if one adopts the point of view that the reality of physical magnitudes is generally unsharp
and seeks an explanation for an objective actualization of properties during measurement, the insolubility 
theorem is most naturally read as entailing the necessity of a modification of the dynamical law of quantum 
mechanics, given that in any measurement-like interaction between the object system and apparatus there is 
the potential that an indicator outcome is actualized. It suggests that there is room for a modification of the Schr\"odinger 
equation so as to allow for the {\em spontaneous collapse} or {\em dynamical reduction} of quantum states, for example,
along the lines proposed by Pearle \cite{Pearle76} and Ghirardi, Rimini, and Weber \cite{GRW86}. 

\subsection{Unsharp objectification}

The question has been raised as to whether the generalized notion of unsharp reality  underwrites a process of
{\em unsharp objectification} that would allow one to avoid the conclusion of the insolubility theorem without 
the need for a modified dynamical law. A partial result has been obtained to the negative \cite{Busch98}, but this
is still based on the requirement of the---possibly unsharp---pointer observable admitting definite values.
The idea of unsharp objectification is based on the description of a pointer observable as an {\em intrinsically
unsharp} observable. Such observables, unlike sharp observables, do not admit outcomes that can occur with certainty. 
Technically speaking, the positive operators (other than the identity $I$) in the range of an intrinsically unsharp observable do not have eigenvalue 1.

If a pointer observable $Z$ (defined as a POM on the (Borel) subsets of $\mathbb R$, say) is intrinsically unsharp, 
so that its effects ($\ne I$) do not admit probabilities equal to one, then the measured observable $E$  defined 
according to the probability reproducibility condition is {\em also} an intrinsically unsharp observable given 
as a POM on $\mathbb R$. Thus, although a general proof is still outstanding, we would expect that a unitary 
measurement coupling cannot turn an element of unsharp reality with a potentiality (probability) of (say) 
1/2  into an element of approximate reality with probability close to 1.

It is interesting to note here Peter Mittelstaedt's comment on unsharp objectification. In \cite{Mitt07} he writes:
\begin{quote}
``It is well known, that within the framework of quantum mechanics it is not possible after a unitary premeasurement 
to attribute sharp or unsharp values to the pointer observable (Busch/Lahti/Mittelstaedt 1996; Mittelstaedt 1998 and Busch 1998). 
Even if merely unsharp values are attributed to the pointer, these values turn out to be not strictly reliable.$^1$ For macroscopic 
quantities, however, this unreliability is practically negligible, and hence it was never observed. On the other side, we should keep 
in mind that in quantum physics the requirement of objective and reliable pointer values - the pointer objectification postulate - is 
merely a reminiscence to classical physics. However, classical physics is based at least partially on ontological hypotheses 
without any rational or empirical justification. Hence, there is in principle no reason to maintain the requirement of objectification\ldots

$^1$ This means that `\ldots one cannot claim with certainty that the reading one means to have taken is reproducible on a 
second look at the pointer' (Busch 1998, p. 246)."
\end{quote}

\noindent
This argument seems to suggest that the notion of objective properties is based on ontological premises made tacitly in 
classical physics that were found empirically untenable and invalidated in the realm of quantum mechanics. If one assumes that 
Quantum Mechanics supersedes Classical Mechanics, it would then follow that it is the requirement of objectification that is 
to be abandoned. This begs the question of what measurements are about if they cannot be regarded as providing definite 
outcomes in some sense.

However, there is an alternative possibility to the universality of Quantum Mechanics that has been put forth by a 
number of physicists, prominently including G\"unther Ludwig: it may eventually turn out to be the case that both 
Quantum Mechanics and Classical Mechanics are limiting cases of a more general and more comprehensive 
physical theory that assigns different domains of validity to these two theories. Such an extended framework 
would leave room for a notion of objective reality as it underlies the objectification requirement.

We mention two examples of theory or model constructions wherein this scenario has been schematized.
The first example is given by the $C^*$-algebraic approach to the formulation of physical theories, which 
provides a framework where the structure of the algebra of observables can either be Abelian (representing 
the classical case), or irreducible (as in standard Quantum Mechanics), or intermediate. In the last case the center 
of the algebra is neither trivial nor does it comprise the whole algebra. This corresponds to a quantum system with 
superselection rules, which may arise in relativistic quantum field  theories or in quantum theories of macroscopic 
systems. In this approach there is thus a rigid distinction between cases with or without superselection rules, and 
the emergence of classical properties in the modeling of larger and larger quantum systems is somewhat 
discontinuous, as it requires a limit to a genuinely infinite number of degrees of freedom. Moreover, the known
rigorous model solutions to the measurement problem, which describe pointers as macroscopic observables,
entail that the actualization of definite outcomes requires infinite time (for a review of this approach to the
measurement problem, see \cite{QTM}).

The second example that allows hybrid  classical-and-quantum  behavior are the alterations of Quantum 
Mechanics with modified  Schr\"odinger equations, such that spontaneous collapse happens at a rate that is 
related to the size of the system. In such approaches, the transition from quantum to classical features 
seems more continuous than in the former approach. The criticism has been put forth that despite
the fact that spontaneous collapse happens practically instantaneously, it is never actually complete, in the 
sense that it would leave the system in a single component  state of its initial superposition. The distribution 
over the different component states is rather sharply peaked but never strictly concentrated on a single state. 
This is the so-called {\em tail problem} \cite{Ghirardi02,Pearle07,Pearle09}.

It seems that the tail problem can be accounted for naturally as an instance of the notion of
approximate reality. Furthermore, it appears that the discontinuity problem for classical properties in the algebraic approach can 
at least be alleviated by giving due account of the nature of a supposedly classical pointer observable. In fact, prototypical pointer 
observables in a quantum measurement are the locations of pointer needles, and thus of systems with simultaneously rather well-defined 
(on a macroscopic scale) positions and momenta. As pointed out in Subsection \ref{subsec:clqm}, the natural representation of such 
observables within the framework of Quantum Mechanics is in terms of (covariant) phase space  observables with intrinsic 
inaccuracies that are macroscopically large compared to Planck's constant but that can still be considered small
on the scale of macroscopic measurement accuracies \cite{OQP}.

\subsection{Summary: emergence of classicality} 

From the perspective of a position that takes Quantum Mechanics to be universal, a resolution of the quantum--classical contrast as
reflected in the above four problems should be envisaged as follows. First, {\em quantization} should be recast as a procedure to 
formulate the quantum theory of a type of object
{\em intrinsically}, without recourse to a classical physical background. This will presumably have to comprise a quantum account
of spacetime. The {\em dequantization} (hidden variables) problem just highlights the fundamentally nonclassical nature of quantum 
mechanics and thus demonstrates that the {\em classical limit} can only be expected to account for an {\em approximate} emergent 
classical behavior of macroscopic objects and large-scale spacetime structure. 

Finally, if in addition to the universality of Quantum Mechanics the (unsharp) objectification of measurement outcomes is maintained,
the insolubility theorem suggests that this can only be realized by accepting a change of the dynamical postulate so as to allow
for spontaneous collapse of the quantum state. 

If the notion of state collapse as a real, stochastic physical process is considered in the context of localization experiments, one is
forced to address the nonlocal nature of the collapse: the fact that a quantum object has been localized in one space region entails
that the probability of its being localizable in another region has instantly become zero; the wavefunction (state in the position representation)
will have collapsed to zero everywhere except in the region of localization. In the language of potentialities or their quantification as degrees
of reality, this means that the system's capacity to trigger a localization event has instantaneously changed to zero outside the localization region. 
As is known from the no-signaling theorem, this collapse does not give rise to superluminal causal influences that could be used as signals. 
Still, this abrupt change of the spatial distribution of the potentiality of localization appears to call for a causal account. 
Such an account could conceivably be obtained in a quantum theory that provides a description of the system {\em and} spacetime as  
quantum dynamical entities. Then the so-called spontaneous collapse could be construed as induced by dynamical
features of spacetime.

\section{Elements of unsharp  quantum reality and their evolution}\label{sec:uqr}

In this section, we put the realist interpretation of Quantum Mechanics in terms of elements of (generally) unsharp reality
to work in order to show how it affords a richer description of quantum systems than the standard account in terms of 
sharp realities alone. We begin with a historical commentary.

John Bell believed that the idea that quantum systems might be ``unspeakable,'' in that one cannot
properly make statements about the properties of quantum systems but only about apparatus
that measure them, was an error of the ``founding fathers'' of quantum theory \cite[p.~171]{BellSpeak}.
Instead of accepting such a restriction, realist interpretations of quantum states (in the physical sense of
the term {\em realist}) consider the values precisely 
ascribed by pure quantum states as actual, and these actual properties constitute the {\em ontic state}
of the system; any change in a pure state (be it through free evolution or measurement) is associated 
with a corresponding change in the ontic state. 

Contrary to Bell's contention, such a realist view was in fact held  by one of the founding fathers of quantum theory, Werner Heisenberg,
as laid down in his later work on the foundations of the theory \cite{H1}. 
He considers the `probability function' (pure or mixed quantum state) as comprising objective and subjective 
elements. The former are ``statements about possibilities or better tendencies (`potentia' in Aristotelian 
philosophy)" \cite[p.~53]{H1},  
whereas the latter represent the observers' limited knowledge; this subjective element  (lack of knowledge) is said to be negligible 
in the pure case. In particular, Heisenberg puts great emphasis on the ``notion that a course of events in itself is not determined
by necessity but that the possibility or rather the `tendency' towards a course of events possesses itself a kind of reality---a certain
intermediate level of reality midway between the massive reality of matter and the mental reality of an idea or picture..."
\cite[p.~140]{H2} (our translation). 
He notes that this concept of possibility is given in quantum theory in the form of probability. 

Heisenberg thus goes beyond the usual interpretation of the pure quantum state $\psi$ as the catalogue of all 
actual properties---those with probability equal to one---of an individual system in that he considers $\psi$ as the
catalogue of the potentialities of all possible (sharp) properties $Q$ of the system, quantified by the probabilities
$p_\psi(Q)=\langle\psi |Q\psi\rangle$.  The notion of actualization of potentialities then 
makes it possible for one to ``say that the transition from the `possible' to the `actual' takes place as soon as the
interaction between the object and the measuring device, and thereby with the rest of the world, has come into
play; it is not connected with the act of registration of the result in the mind of the observer.''
\cite[pp.~54-55]{H1}.

In the more recent work in the philosophy of physics, potentiality has been further articulated by Shimony as a modality of 
existence of physical systems that confers an intermediate status to properties between that of bare logical possibility and 
full actuality \cite{Shimony}. As Shimony points out, the existence of such an intermediate modality is in fact {\em required}
by the non-local correlations violating Bell's inequality \cite[Vol.~II p.~179]{Shimony}. This connection is in accordance with 
the fact that, formally, entanglement is an instance of the superposition of  mutually orthogonal product states, and thus 
gives rise to quantum indeterminacy.

Here we propose to add the interpretation of potentiality as a {\em causal agency}:  the tendency or propensity of 
the property or effect in question to actualize a value or an indication of itself in a measurement device constitutes a
{\em probabilistic cause} for this actualization. Insofar as reality qua actuality is the capacity to act, that is, to cause 
indicative events, the potentiality of a quantum event then becomes  its unsharp reality; the corresponding 
quantum probability gives its degree of reality, that is, its quantified limited capacity to cause an indicative event.\footnote{In current
research in the philosophy of physics a case is being made for the causal power of dispositions, or potentialities; this is laid out 
in the context of the quantum measurement problem in a contribution to this Festschrift by M.~Esfeld \cite{Esfeld2010}.}

To further explicate this extended single-case interpretation of quantum mechanical probabilities, we take up the  
observation, noted earlier, that the pure state $\psi$ of an isolated quantum system, represented as a projection $P=P_\psi$, 
can be viewed as a point in this system's phase space $\Gamma_q$ that evolves deterministically along a 
trajectory given by the Schr\"odinger equation. This time development gives rise to a continuous evolution of the
potentialities, that is, degrees of reality $f_E(P)={\rm tr}[P E]=\langle\psi |E\psi\rangle$ of all the effects $E$ of the system, which thereby are all
simultaneously real to a degree (actualization tendency) given by $f_E(P)$.

In addition to such attribution of degrees of reality,
the use of POMs enables the simultaneous measurement of approximate values of noncommuting properties. In fact,
the joint measurement of properties for which sharp observables are incompatible is possible: Although
non-commuting sharp (standard) observables are never jointly measurable, they are capable of 
approximation by a pair of related unsharp observables, represented by suitable POMs, which {\em will be} jointly measurable.
Moreover, similar to L\"uders measurements, which are perfectly repeatable, it is possible to perform joint approximate measurements
that have approximate repeatability properties, leading to increased degrees of reality of the measured properties.

Thus, let us take quantum probability as quantifying the tendency 
to actualize properties in the ontic sense for the individual, as opposed to the merely epistemic sense. 
That is, let us take the expression $p_\psi^E(X)= 
\langle\psi |E(X)\psi\rangle$, for effect $E(X)$ associated with the value set $X$, as providing the likelihood of the actualization 
of the potential property, whether sharp or unsharp, when measured on a system prepared in pure state 
$\psi$. 
In the standard  account, the measured system interacts with a probe system, 
resulting in an entangled state for the total system. We can assume a similar scheme for approximate joint
measurement of both position and momentum since the existence of such schemes is warranted by the existence of 
POMs that represent joint observables for momentum and position (as reviewed in \cite{BHL2007}). 

If the couplings between the object system and two appropriate measurement  probes are activated 
simultaneously, the resulting measurement scheme constitutes a joint approximate measurement of both 
position, $q$, and momentum, $p$. No matter how small  the position and momentum imprecisions if  
measurements are made of them separately, the joint coupling of both probes results in a re-adjustment of the 
individual measurement imprecisions in such a way that they satisfy an uncertainty relation. It is possible to
choose measurement couplings such that the joint measurement is approximately repeatable \cite{OQP}. 
Thus, upon obtaining a joint reading of $q$ and $p$, the quantum particle will be found afterwards in a state in which 
position and momentum are unsharply localised at the phase space point $(q,p)$: the centers and widths of 
the associated position and momentum distributions are equal to the values $q, p$ and the measurement 
imprecisions $\delta q,\delta p$, respectively. 

The example of approximately repeatable phase space measurement provides a theoretical underpinning of the
reality of bubble chamber or cloud chamber tracks of elementary particles. It also corroborates Heisenberg's famous
dictum, ``I believe that the existence of the classical `path' can be pregnantly formulated as follows: The `path' comes 
into existence only when we observe it." \cite[p.~185]{Heis27}.
We are interested in observing these tracks because they are probes for what is going on in the high energy collisions 
or decays in which these particles are produced. One may even say that the approximate repeatability is necessary 
because without it one cannot reconstruct the energy-momentum balance of the processes. 

These considerations show that the possibility of simultaneously attributing fuzzy values to noncommuting quantities 
in a given state is underwritten by the quantum theory of measurement, according to which these values have likelihoods 
of actualization (thus degrees of reality) that can be tested simultaneously in appropriate measurements.
An indeterminacy interpretation of quantum uncertainties thus establishes consistency between the possibilities of 
state definition, that is, the preparation of noncommuting sets of effects and the possibilities of determination, that is,
the joint measurement of these quantities.

When a measurement is performed, there is a stochastic change of the states of both the measured 
system and the measuring system, whether either a sharp or an unsharp observable is measured. 
When a {\em repeatable} measurement is made of a sharp observable of the measured system, the
properties associated with sharp observables compatible with that observable become actual and 
definite and those associated with incompatible sharp observables become indefinite and potential. 
When a measurement is made of an {\em unsharp} observable, the property's value is fuzzy but 
can become less indeterminate and property values associated with compatible unsharp observables 
remain fuzzy and may become either more or less indeterminate. 

To summarize, given the POM formalism, the ontic potentiality approach gains further reach than is 
the case given only the PVMs; one is able to provide further explanations of the behavior of quantum systems, 
because one can make use of unsharp observables to explain the results of joint measurements of physical 
properties traditionally thought of as incompatible and unspeakable. In particular, better predictions of 
subsequent measurements can be provided by the updated probabilities corresponding to the observed 
outcomes, whether the observable to be measured is sharp or unsharp, so long as the corresponding 
uncertainty product is reduced. 

In addition to accommodating the restrictions on the joint determination of properties of quantum systems, 
the use of the POM formalism under the above ontic potentiality interpretation allows one conceptually to 
ground quantum probability in (probabilistic) causality and to provide explanations for the statistics of outcomes of both 
sharp and unsharp measurements of individual systems, rather than leaving many situations ``unspeakable.''

Depending on an experimenter's interests, he can choose to make either a sharp measurement and determine 
a single property of a complementary pair or he can make an optimal joint unsharp of a pair of corresponding 
indeterminate properties. Although, when joint approximate measurements are made of both of a pair of physical 
magnitudes, say position and momentum, there are only fuzzy values and potentially sharp properties, both properties 
do exist and can be spoken of. 

Quantum fuzziness is not a reflection of some theoretical deficiency in this interpretation but rather is naturally 
related to quantum probability which is grounded in causality. Although the sharp properties are not actualized 
during approximate, unsharp measurements, the likelihoods for the value ranges corresponding to the outcomes 
that are found can increase. In this way, a clear and  consistent meaning is given to (generalized) quantum 
observables, namely, as unsharp properties always consistently jointly attributable to individual quantum systems, 
where these properties and their degrees of reality have an evolution that is explicable and can be used to 
provide physical explanations.

\vspace{12pt}
\noindent {\sc Acknowledgement.}\\
We would like to thank Pekka Lahti and Leon Loveridge for their critical reading and valuable comments on draft versions of this paper.


\end{document}